\documentclass[manuscript]{acmart}

\AtBeginDocument{%
  \providecommand\BibTeX{{%
    \normalfont B\kern-0.5em{\scshape i\kern-0.25em b}\kern-0.8em\TeX}}}

\setcopyright{acmcopyright}
\copyrightyear{2024}
\acmYear{2024}
\acmDOI{XXXXXXX.XXXXXXX}

\begin{document}


\title{Human-AI Collaboration in Thematic Analysis using ChatGPT: A User Study and Design Recommendations}


\author{Lixiang Yan}
\affiliation{%
  \institution{Monash University}
  \country{Australia}
}

\author{Vanessa Echeverria}
\affiliation{%
  \institution{Monash University}
  \country{Australia}
}

\author{Gloria Fernandez Nieto}
\affiliation{%
  \institution{Monash University}
  \country{Australia}
}

\author{Yueqiao Jin}
\affiliation{%
  \institution{Monash University}
  \country{Australia}
}

\author{Zachari Swiecki}
\affiliation{%
  \institution{Monash University}
  \country{Australia}
}

\author{Linxuan Zhao}
\affiliation{%
  \institution{Monash University}
  \country{Australia}
}

\author{Dragan Gašević}
\affiliation{%
  \institution{Monash University}
  \country{Australia}
}

\author{Roberto Martinez-Maldonado}
\affiliation{%
  \institution{Monash University}
  \country{Australia}
}

\renewcommand{\shortauthors}{Yan et al.}

\begin{abstract}

Generative artificial intelligence (GenAI) offers promising potential for advancing human-AI collaboration in qualitative research. However, existing works focused on conventional machine-learning and pattern-based AI systems, and little is known about how researchers interact with GenAI in qualitative research. This work delves into researchers' perceptions of their collaboration with GenAI, specifically ChatGPT. Through a user study involving ten qualitative researchers, we found ChatGPT to be a valuable collaborator for thematic analysis, enhancing coding efficiency, aiding initial data exploration, offering granular quantitative insights, and assisting comprehension for non-native speakers and non-experts. Yet, concerns about its trustworthiness and accuracy, reliability and consistency, limited contextual understanding, and broader acceptance within the research community persist. We contribute five actionable design recommendations to foster effective human-AI collaboration. These include incorporating transparent explanatory mechanisms, enhancing interface and integration capabilities, prioritising contextual understanding and customisation, embedding human-AI feedback loops and iterative functionality, and strengthening trust through validation mechanisms. 


\end{abstract}

\ccsdesc[500]{Human-centered computing~User studies}
\ccsdesc[500]{Human-centered computing~Collaborative interaction}

\keywords{Human-AI Collaboration, Thematic Analysis, Qualitative Research, Generative Artificial Intelligence, ChatGPT}
\maketitle

\section{Introduction}

Human-AI collaboration in qualitative research is gaining traction within the human-computer interaction (HCI) community \cite{amershi2019guidelines, wang2020from, yang2020reexamining}. This collaboration reflects the paradigm shift from complete automation by AI to a more collaborative role that AI plays in human workflows, where humans and AI work together to achieve a shared goal \cite{harper2019role, wang2020from, lai2022humanai}. Such collaborations present opportunities for both enhancing research efficiency and deepening analytical insights \cite{lennon2021developing}. Specifically, in the domain of thematic analysis, various prototype systems have been introduced, demonstrating different automation levels through the integration of pattern-based and machine-learning methods \cite{rietz2021cody, gebreegziabher2023patat, gao2023coAIcoder}. While these systems have demonstrated efficacy in enhancing coding quality and improving inter-coder reliability \cite{rietz2021cody, gebreegziabher2023patat}, they were difficult to scale with diverse linguistic features due to limited pattern rules or model training requirements, which is one of the unique features of qualitative data \cite{braun2021qualitative}.

Recently, the emergence of generative artificial intelligence (GenAI), exemplified by ChatGPT and other state-of-the-art large language models, underscores the capabilities of advanced AI to process and analyze unstructured text-based qualitative data without additional model training or human intervention \cite{van2023chatgpt, li2023reflective, yan2023llms}. Such technologies, equipped with extensive pre-trained knowledge, have the capacity to identify data patterns that may not be immediately apparent to researchers \cite{zambrano2023ncoder, gao2023collabcoder}. Yet, the implications of integrating GenAI into the qualitative analysis workflow, inherently iterative and human-centred, remain underexplored. Additionally, little is known about how to optimise researchers' collaborative experience with GenAI in thematic analysis as such systems offer a high degree of user autonomy through natural language interaction \cite{pavlik2023collaborating, sallam2023ChatGPT}, which is substantially different from prior machine-learning and pattern-based systems. In light of this, our user study delves into the dynamics of how qualitative researchers engage with ChatGPT in the context of thematic analysis. The primary contribution of our research includes five actionable design recommendations synthesised from the benefits and challenges that qualitative researchers experienced when collaborating with ChatGPT for thematic analysis. These recommendations can offer guidance to HCI scholars in designing and developing applications that not only optimise the process of human-AI collaboration in thematic analysis but also fortify user trust and confidence in the technology.

\section{Related Work}

\subsection{Thematic Analysis in Qualitative Research}

Thematic analysis is a qualitative research method that involves uncovering and analyzing themes or patterns within qualitative data, such as interview transcripts and open-ended responses \cite{braun2006using}. As conceptualized by Braun and Clarke \cite{braun2021qualitative}, depending on the epistemological assumptions, this multifaceted method can be undertaken deductively, where researchers analyze and interpret data through the lens of existing research and theory, or inductively, where the analysis is grounded in the data. Regardless of the approach, it is an iterative method, where researchers need to go over the data multiple times to gain an in-depth understanding of the data, articulate, refine, and confirm the themes or patterns that they have identified \cite{boyatzis1998transforming}. Consequently, thematic analysis is a time-demanding and skill-intensive task that may prove challenging to scale with a large corpus of qualitative data \cite{marathe2018semi,chen2018ambiguity}.

\subsection{Human-AI Collaboration in Thematic Analysis}

Research on human-AI collaboration in thematic analysis has produced several prototype systems to support automated or semi-automated coding, including text classification, topic modelling, and annotations of qualitative data \cite{rietz2021cody, gebreegziabher2023patat, lennon2021developing, goudjil2018novel}. These systems have leveraged a mix of techniques, ranging from machine-learning techniques for text classification \cite{blackwell2018computer, rietz2021cody, yan2014semi, goudjil2018novel} to interactive topic modelling algorithms \cite{bakharia2016interactive, baumer2017comparing, leeson2019natural, lennon2021developing, hong2022scholastic} and adaptive pattern-based rules \cite{crowston2010machine, gebreegziabher2023patat, kaufmann2020supporting, marathe2018semi, rietz2021cody}. Systems based on machine-learning techniques like logistic regression \cite{rietz2021cody} and support vector machine models \cite{yan2014semi} have demonstrated potential in automating large-scale text data annotation. Topic modelling algorithms, especially variations of Latent Dirichlet Allocation, have shown performance comparable to open coding in identifying underlying topics in qualitative data \cite{bakharia2016interactive,hong2022scholastic,lennon2021developing}. These techniques, however, raise concerns about transparency in their decision-making, which may challenge the primary objective of thematic analysis: gaining an in-depth understanding of qualitative data \cite{chen2018ambiguity, marathe2018semi}. In contrast, adaptive pattern-based systems such as Cody \cite{rietz2021cody} and PaTAT \cite{gebreegziabher2023patat} adopted user-interpretable patterns to enhance user understanding and trust. Yet, their reliance on pattern-based rules might constrain their analysis of data with varied linguistic features \cite{gebreegziabher2023patat}. Additionally, these pattern-based systems may also be unsustainable considering the nuances in the linguistic and semantic features of different languages \cite{eisenschlos2019multifit}.

\subsection{Generative Artificial Intelligence}

GenAI has witnessed significant progress in recent years, providing a foundation for developing solutions capable of generating human-like contents \cite{sallam2023ChatGPT}. The core idea behind GenAI is to train models on vast amounts of data, enabling them to produce original content that closely mirrors human language patterns \cite{pavlik2023collaborating,van2023chatgpt}. Among the notable advancements in this field are large language models like ChatGPT, which have been trained on diverse textual data sources, allowing them to comprehend and generate a wide range of language structures \cite{sallam2023ChatGPT,yan2023llms}. Beyond text generation, these models are also capable of understanding context, offering explanations, and assisting in tasks such as thematic analysis \cite{zambrano2023ncoder,amarasinghe2023generative}. The vast knowledge encoded in these models, coupled with their ability to generalize from training data, positions them as potential collaborative partners in qualitative research \cite{gao2023coAIcoder}. Their ability to capture nuanced language structures and provide contextually relevant content has the potential to support qualitative researchers in the labour-intensive process of thematic analysis \cite{zambrano2023ncoder}. However, while GenAI exhibits promise, it also raises questions about its suitability and reliability in qualitative research contexts \cite{bang2023multitask, reiss2023testing}. For instance, the reliability of the decisions made by these models and their trustworthiness remain topics of discussion \cite{van2023chatgpt}, especially when used in thematic analysis, where understanding and transparency are paramount \cite{braun2021qualitative, gebreegziabher2023patat}. Thus, the interplay between GenAI technologies and qualitative researchers requires further investigation to uncover collaborative potential and areas of concern following a human-centred AI approach \cite{shneiderman2020human, shneiderman2020bridging}.

\subsection{Research Questions}

The current user study aims to address the aforementioned gaps by delving into qualitative researchers' perceptions of their collaboration with GenAI, specifically ChatGPT, and highlighting actionable design recommendations to foster effective human-AI collaboration with this novel technology. Specifically, the following research questions were addressed.

\begin{itemize}
    \item RQ1: In the context of thematic analysis, what \textit{opportunities} do qualitative researchers identify when collaborating with ChatGPT?
    \item RQ2: While collaborating with ChatGPT for thematic analysis, what \textit{challenges} do qualitative researchers encounter?
\end{itemize}

\section{Methods}

\subsection{Participants}

We conducted a user study to investigate the experience of qualitative researchers when collaborating with ChatGPT to conduct thematic analysis, with a specific focus on uncovering the opportunities and challenges that they perceived during this human-AI collaboration process. We recruited participants following a criterion-based sampling via a faculty and a personal mailing list. Participants were doctoral students or full-time academics with at least one year of experience in qualitative research and had conducted thematic analysis, either inductively or deductively, themselves. Three full-time academics (P1--P3) and seven doctoral students (P4--P10) with different years of experience (M = 4.0, SD = 3.3; four females) in thematic analysis participated in this study. Ethics approval was obtained from Anonymous University (Project ID: Anonymized).

\subsection{Study Design}

We adopted a contextual inquiry approach to guide the study design \cite{karen2017contextual}. This user-centred approach has been used in prior HCI research to uncover enriched insights regarding users' authentic experiences and needs of specific tasks in their natural environment or context \cite{rietz2021cody}. Each session was around one hour and contained two main sections: 

\begin{enumerate}
    \item \textbf{Thematic Analysis Exercise} (20 mins): Participants were first asked to manually perform inductive thematic analysis on a short transcript containing 15 utterances. Following Braun and Clarke's \cite{braun2021qualitative} guidelines, the coding process was open and organic without any pre-existing coding framework and the final outcome was the themes that participants identified. This practical exercise aimed to provide participants with contextual knowledge of the transcript, which was required for them to evaluate the outputs from ChatGPT in the second task.
    \item \textbf{Collaboration with ChatGPT} (40 mins): After analyzing the short transcript, participants were given an introduction to ChatGPT (5 mins). We then provided participants with an extended version of the short transcript, containing 15 additional utterances (30 utterances in total). Participants were given access to ChatGPT (GPT-4 Default mode) and were asked to collaborate with ChatGPT to perform thematic analysis on the extended transcript while conducting an in-situ evaluation of the outputs with the think-aloud method (15 mins). Finally, we evaluated participants' experience of collaborating with ChatGPT for thematic analysis through a semi-structured interview (20 mins). Specifically, participants were asked to reflect on:
    \begin{itemize}
        \item Accuracy of the AI-generated results.
        \item Trustworthiness of the AI-generated results.
        \item Helpfulness of ChatGPT for thematic analysis. 
    \end{itemize}
    Participants were also asked about how the current interface and functionality of ChatGPT can be improved to better accommodate their research needs for thematic analysis. All sessions were conducted remotely via Zoom, during which we captured audio, shared screen visuals, and archived the chat history between participants and ChatGPT.
\end{enumerate}

\subsection{Analysis}

Using Otter AI (otter.ai), we transcribed each session's audio recording. Addressing our research questions, the primary author conducted a reflexive thematic analysis \cite{braun2021qualitative} of these transcripts in conjunction with each participant's ChatGPT chat history. This initial analysis was augmented by discussions with a co-author, facilitating the iterative refinement of emergent themes related to participants' perceived opportunities (RQ1) and challenges (RQ2) when collaborating with ChatGPT to conduct thematic analysis. We commenced our analysis on an individual participant basis, extracting preliminary codes pertinent to each RQ. Subsequently, we aggregated these preliminary codes into overarching themes for every RQ. The goal of this analysis was to comprehend the nature of participants' interactions with ChatGPT and inform design recommendations for future HCI studies.

\section{Findings and Discussion}

In this section, we first reported the opportunities and challenges participants experienced when collaborating with ChatGPT to conduct thematic analysis. Based on these findings, we synthesised five actionable design recommendations to foster effective human-AI collaboration.

\subsection{Human-AI Collaboration Opportunities (RQ1)}

Participants reported several multifaceted opportunities for human-AI collaboration, particularly leveraging tools like ChatGPT, in thematic analysis. 

\subsubsection{Efficiency in Processing and Analysis} The efficiency of ChatGPT in processing large datasets and distilling themes was a salient observation. Such proficiency offers a promising avenue for the swift analysis of unstructured qualitative data. P1 articulated: \textit{"ChatGPT provides an efficient method to identify key themes from qualitative data."} This sentiment was echoed by P7, who observed: \textit{"For this task, I think, yeah, it definitely can save my time. Yeah, to identify emerging themes."} P10 went a step further, emphasizing the technology's superiority in certain aspects of analysis: \textit{"I think it's good at doing this kind of summarization task. Even better than human."} The advantages of ChatGPT become even more pronounced when handling voluminous datasets. P8 remarked: \textit{"Especially if I have a lot of data and I really don't have time commitment or to analyze it by myself."} Beyond its efficiency, ChatGPT's capabilities in transforming unstructured datasets into structured formats were also recognized. P2 commented: \textit{"I was impressed at its ability to, at least for some of the data, format it in a way that I asked it to right with, like the table and stuff."}

\subsubsection{Facilitation of Initial Exploration} Participants recognized the instrumental value of ChatGPT during the preliminary stages of both inductive and deductive analytical processes. They posited that ChatGPT could offer a foundational coding framework that would be particularly beneficial during the initial stages of code generation, exploration, and ideation. P2, for instance, highlighted the tool's capacity for summation and ideation by stating: \textit{"I have identified that ChatGPT is very useful to summarize general topics, so I will use it for inspiration, and to identify codes in the process."} In the context of the inductive coding approach, P8 underscored ChatGPT's potential for exploration: \textit{"So, for instance, we were talking about the inductive way of coding, I will ask ChatGPT, can you give me the list of the most prominent themes which are presented in the piece of text."} Extending on this, P8 also suggested that the themes generated by ChatGPT could set the stage for developing an initial coding framework for deductive analysis: \textit{"When we're doing deductive analysis, it [ChatGPT] will give the list of themes, and if we ask ChatGPT, okay, can you try to find instances [utterance] of those themes. It [ChatGPT] also will work fine."} This potential for initiating the coding process in deductive analysis was also evidenced by P9, who highlighted the potential for coding interview transcripts: \textit{"I will have multiple interview transcripts, and I will code maybe one or two transcripts, and then I will probably pass it to ChatGPT, and maybe it will learn the patterns and help me to code the transcripts."} 

\subsubsection{Quantitative Insights and Detailed Metrics} Beyond assisting with the initial exploration stages, ChatGPT's ability to offer quantitative insights and detailed metrics like coverage is appreciated. For example, P1 highlighted its strength in providing a detailed breakdown of theme coverage through prompting: \textit{"I think that's really useful getting all these metrics of coverage."} Such metrics can help researchers understand the extent to which a particular theme or topic is represented within the dataset. P2 echoed this sentiment, emphasizing ChatGPT's capacity to shed light on \textit{"the frequency or popularity of certain themes or topics."} This capability can be instrumental in discerning recurrent themes, thereby paving the way for deeper qualitative investigations, such as identifying sub-themes. Meanwhile, P7 also saw the potential of ChatGPT in analyzing coded data: \textit{"But after coding, we still need to do some analysis, right? We are not only going to do some frequency tables. If it can help me with analyzing coded data to identify some patterns or some preliminary analysis, it will be good."}

\subsubsection{Support Language Comprehension} Participants also appreciated ChatGPT's assistance in language comprehension, especially for non-native English speakers. For example, P4 reflected that: \textit{"Not only about time-wise, but also sometimes because, for example, for me, English is not my first language. Sometimes the interviewees may say something too casual [or] too verbal that I couldn't understand, [but] I think ChatGPT could.}" P10 pointed out ChatGPT's value in understanding language usage in complex or unfamiliar domains: \textit{"I think a human has some limitation of knowledge. For example, maybe I don't understand when talking about politics, but ChatGPT can understand it better."} 

\subsubsection{Discussion of RQ1} The findings on the efficiency and efficacy of ChatGPT in supporting the initial exploration of qualitative data resonated with prior works \cite{zambrano2023ncoder, goudjil2018novel, rietz2021cody, gebreegziabher2023patat}, illustrating the collaborative synergy between researchers and AI, where they both contribute to the process, instead of completely outsourcing the task to AI. The appreciation for quantitative insights and detailed metrics further strengthened such collaborations where researchers can leverage these additional insights offered by AI to conduct further in-depth analysis. This finding also reinforced the need for human-centred AI where the goal of well-designed AI systems should be supporting humans instead of replacing them \cite{shneiderman2020human, shneiderman2020bridging}. Apart from these findings, we also identified a unique benefit of GenAI, that is, supporting the language comprehension of non-native speakers and non-experts. Such a novel finding is unseen in prior works on machine-learning and pattern-based AI systems \cite{rietz2021cody, gebreegziabher2023patat, bakharia2016interactive, kaufmann2020supporting}. This finding further highlights the augmentation of human capabilities that human-AI collaboration enables, especially in skill-intensive tasks like thematic analysis, where experience and expertise are essential for ensuring quality results \cite{boyatzis1998transforming, braun2021qualitative}.

\subsection{Human-AI Collaboration Challenges (RQ2)}

In exploring the challenges of human-AI collaboration for thematic analysis, several consistent themes emerged across participants' experiences. 

\subsubsection{Trustworthiness and Accuracy Concerns} Participants consistently expressed concerns about the trustworthiness and accuracy of ChatGPT's outputs. They were sceptical of the results produced and often felt the need to manually verify them. P2 articulated this scepticism: \textit{"If I notice something is not very well performed, I will stop trusting the technology, and I will maybe have to go on my own and check manually."} This sentiment was resonated by P4, who expressed a balanced stance: \textit{"I have mutual experience, a neutral perception on the extent I trust ChatGPT."} Moreover, P6 underscored the indispensable role of human oversight: \textit{"So it needs to be checked by a human before it can be used for any research."} Interestingly, while participants generally acknowledged ChatGPT's proficiency in certain tasks like summarization, doubts emerged when it came to numerical data. As P5 observed: \textit{"It has a good side on the summarizing and like integrating codes like that. I trust that part. But about the coverage and like any number, basically the numbers, I do not trust it."} A point of contention arose around the potential for ChatGPT to generate hallucinated results. P8 cautioned: \textit{"First of all is that, for instance, it might be inaccurate in the sense that the codes might be made up."} Yet, P10 offered a more nuanced view, suggesting that while ChatGPT might occasionally deviate from accuracy in some contexts, thematic analysis might not be one of them: \textit{"When it hallucinates, it's usually under the context when I ask for some specific literature, but in this task [thematic analysis] we give it [ChatGPT] the material and let it summarise, seems like an easier task."} These varied perspectives accentuated P9's call for validation works: \textit{"So one thing, we need some literature to show the ChatGPT can actually be used for thematic analysis."}

\subsubsection{Reliability and Consistency Issues} Apart from participants' inherent trust in the technology, the challenge of obtaining consistent results from ChatGPT was another common concern. For example, P3 expressed concerns about the temporal consistency of the system: \textit{"I'm wondering how consistent ChatGPT is over time, you know, with the same stuff."} This sentiment hints at potential intra-reliability issues, as P3 further explained: \textit{"It [ChatGPT] might need some refinement with itself. There are intra-reliability problems that might be happening."} This concern was echoed by P6, who observed variances in the system's outputs: \textit{"I know sometimes the answers can be quite random and inconsistent, no matter how you prompt the request."} Intriguingly, P6 also noted potential differences in results even when using identical prompts: \textit{"Yeah, sometimes with the same prompt. If you use it for the first time, you get great results. But when you use it for even the same transcript, it can produce different results."} Highlighting the broader implications of this inconsistency, P8 underscored the user's inclination towards more deterministic systems: \textit{"You're more inclined to serve the trust to the software, which is more deterministic."}

\subsubsection{Limited Data Capacity and Contextual Understanding} Several participants pointed out ChatGPT's limitations in terms of data processing capacity and its understanding of context. P3 mentioned: \textit{"It [ChatGPT] has some limitations to the amount of data you can give it."} P9 resonated with this concern: \textit{"From my previous experience with ChatGPT, it has a problem with the length of the token [input text]."} This capacity issue may limit the length of the transcript that ChatGPT can analyze, as P9 elucidated: \textit{"The transcripts for thematic analysis [are] usually very long. And in that case, we can only input one session to make it [ChatGPT] learn."} This segmented input could compromise the quality of outcomes, with P9 noted: \textit{"If it [transcripts] cannot be inputted as one message, from my previous experience, the learning efficiency [of ChatGPT] will be harmed."} Another concern is ChatGPT's ability to infer deeper contexts. P4's critique explained this: \textit{"Sometimes ChatGPT only provides information based on the text, and sometimes the text may not be able to 100\% reflect on the key things that the [participant] wants to express."} Drawing from personal experiences in domain-specific research, P4 underscored the necessity of a contextualized understanding, especially in thematic analysis. For example, P4 explained the following coming from the context of law education: \textit{"I found that if you had experience learning or teaching experience in law, you would have a better understanding about what the [study participants] want to express."} Meanwhile, P7 highlighted the lack of academic focus of ChatGPT: \textit{"I think it only can identify some emerging themes based on the transcripts, but it's not really academic oriented."} The imperative for deep comprehension of existing academic literature was also highlighted by P7: \textit{"I think it [ChatGPT] needs to understand the literature. So I probably need to upload my literature review or at least what articles that I used to create the coding scheme?"} Without such an enriched contextual foundation, there could be a misalignment between humans and AI. As P10 summarized, \textit{"I feel like it's about the weight of importance that ChatGPT has maybe misaligned with researchers."}

\subsubsection{Interface and Integration Challenges} Other than the aforementioned functionality challenges, participants also discussed the difficulties and inefficiencies in interacting with the current interface of ChatGPT for thematic analysis. P3 highlighted the suboptimal experience: \textit{"And obviously, it's not ideal to be interacting with it in the ChatGPT window."} Expanding on this, P3 underscored the system's inefficiencies when adopting an iterative methodology: \textit{"I will probably do [the analysis using ChatGPT] like 2 or 3 more times, but at least in the way that the interface is set up, it's not conducive to me doing it quickly and easily."} On the same note, P9 emphasized the need for better integration and formatting tools: \textit{"Maybe we can also format the input as a table; maybe we can also import the transcription table directly as an input to the AI."} 

\subsubsection{Acceptance within the Research Community} Participants expressed concerns about the adoption and disclosure of human-AI collaboration in qualitative research. A predominant challenge arises from potential scrutiny by reviewers regarding the tool's reliability and the validation process. This is coupled with the broader acceptance of such technologies within the scientific community. P6 highlighted the dilemma: \textit{"We use ChatGPT. We want to disclose it. However, we can imagine there could be challenges from reviewers about reliability and accuracy. And how do you validate the process?"} This participant further commented on the societal perceptions around the use of GenAI technologies like ChatGPT, especially in academic contexts. As P6 explained: \textit{"I think when I publish something, I'll tell everyone. Okay. I used ChatGPT to polish my articles. That just sounds sort of weird."} P1 echoed the sentiment of uncertainty but also expressed optimism about the future adoption of these technologies in research: \textit{"But if it is about the use of AI in research, I think that's a big question, and we don't know what needs to be done at this point."} P1 further articulated hope for a future where the use of GenAI is normalized: \textit{"I guess it will be more accepted as well, and it will be just another method being used like any other methods. So it's not about disclosing if you're using AI or not, but just simply describing the method and the tool that was used."} However, not all views were optimistic. P9 highlighted potential biases within the reviewer community: \textit{"I know some reviewers actually, very hate the AI in doing stuff."}

\subsubsection{Discussion of RQ2} The concerns regarding the trustworthiness, accuracy, reliability, and consistent of AI-generated resonated with previous findings \cite{rietz2021cody, gebreegziabher2023patat, lennon2021developing, hong2022scholastic}, indicating the need to include trust-assuring mechanisms when designing human-AI collaboration. Whereas, the limited data capacity and contextual understanding seem to be a novel challenge for GenAI. While data capacity issues might be resolvable through model advancements and algorithm innovations \cite{chen2023extending}, contextual understanding can only be achieved through effective human-AI collaboration, where humans actively provide and refine the contextual knowledge of AI during an iterative circle \cite{boyatzis1998transforming}. Such a synergy would require designing an interface that prioritises the continued and mutual communication between humans and AI \cite{shneiderman2020human, schmidt2020interactive}. Lastly, an ethical dilemma exists between adopting a human-AI collaborative approach and adequately reporting such usages in academic publications. While this issue remains an ongoing debate, there is an increasing consensus on the need for transparency, disclosure, and clear guidelines to ensure that AI contributions are acknowledged without undermining the authenticity and integrity of the research \cite{dwivedi2023so, van2023chatgpt}. Apart from guidelines, in the context of thematic analysis, empirical evidence is urgently needed to evaluate the validity of the AI-generated results. 
\subsection{Design Recommendations}

Building on our findings, we have synthesised five actionable design recommendations to support future HCI studies in designing and developing AI solutions that streamline the process of human-AI collaboration in thematic analysis while also ensuring user trust and confidence in the technology.

\subsubsection{Incorporate Transparent Explanatory Mechanisms} To increase user confidence and understanding of the technology, it is crucial to design human-AI collaboration solutions with transparent explanatory mechanisms. The AI system should offer detailed rationales behind the generation of specific themes. For instance, by providing relevant data metrics (P2: \textit{"it would be good to see the actual coverage and see it everywhere"}) or highlighting illustrative examples for each theme (P5: \textit{"having quotes... yeah, like for each line, this is the code. And this is the quote"}), users can discern the AI's decision-making process. Incorporating features that allow the AI to show its "thought process" or its basis for conclusions can demystify its operations, ensure that generated themes align with data nuances, and potentially improve its sense of trust.

\subsubsection{Enhance Interface and Integration Capabilities} A human-centred interface redesign is critical. The user interface should be intuitive, allowing for easy navigation and operations such as data uploads and theme suggestions (P3: \textit{"Imagine a step like, I upload my data, and then I click a button and to say like, what do you think the themes are in this?"}). Integration features, like spreadsheet compatibility (P1: \textit{"Maybe a way to submit this spreadsheet... and then the output to be the same spreadsheet"}), can streamline the thematic analysis process, making it more efficient for users. Moreover, considering compatibility with other research tools and platforms can elevate the user experience (P6: \textit{"I wish it could be an all-in-one system that, for example, after this interview... can automatically summarize what we said on Zoom."}), ensuring that human-AI collaboration seamlessly fits into the researcher's workflow.

\subsubsection{Prioritize Contextual Understanding and Customization} The design of human-AI collaboration solutions should prioritize a deep understanding of the data's unique context. Features that allow users to provide contextual cues or specific research backgrounds can enhance the AI's sensitivity to nuances (P4:  \textit{"I prefer to have some kind of tools that I can provide more contextual information to when generating the automatic analysis. Now it's just a transcript."}). Furthermore, customization options, where users can adjust parameters or guide the AI's theme generation based on their specific needs (P5: \textit{"I have the next step, which is to introduce some of my understanding or my knowledge."}), can ensure that the generated themes are rooted in the data's specific context.

\subsubsection{Embed Feedback Loops and Iterative Functionality} A dynamic and collaborative relationship between the user and AI can be fostered by embedding feedback loops within the system. By allowing users to provide feedback on generated themes, and then having AI adjust its outputs accordingly in real-time, the tool can evolve and refine its understanding (P2: \textit{"If I have identified something which is not correct from the tool. If I have some chance to provide additional feedback to the tool to recalculate and to improve."}). Such iterative functionality underscores the idea that thematic analysis is a continuous process (P3: \textit{"Obviously, this could be an iterative process."}), which can benefit immensely from ongoing human-AI collaboration.

\subsubsection{Strengthen Trust through Validation Mechanisms} Building user trust is foundational for effective human-AI collaboration. Designing AI systems with robust validation mechanisms can address concerns regarding their reliability and accuracy. While, the aforementioned transparent explanatory mechanisms, such as coverage metrics, can be an enabler of such validations, more rigorous and empirically-orientated mechanisms are required. For example, features that allow side-by-side comparisons of AI-generated themes with human analysis or that offer manual coding checks can enhance user confidence (P10: \textit{"I need to compare the results generated by ChatGPT and human."}). Such mechanisms could also contribute empirical evidence that is required to convey the validity and reliability of this human-AI collaborative approach to the academic community and external reviewers.

\section{Limitations and Future Works}

This study has several limitations. First, ChatGPT's interface is not optimized for thematic analysis, yet interactions with this GenAI provide insights for future interface design. Second, the interview transcript used has 30 utterances (1,685 tokens) to avoid context length issues (e.g., GPT-4's 8192-token limit), potentially not reflecting challenges with larger datasets. Only two participants (P3, P9) with ChatGPT experience highlighted this issue. Third, the lack of research questions may have limited insights into deductive thematic analysis challenges. Lastly, we identified opportunities and challenges but did not explore the specific processes researchers used during thematic analysis. Future work should design a system informed by this study's findings to support thematic analysis and human-AI collaboration. Evaluating such a system against Cody \cite{rietz2021cody} and PaTAT \cite{gebreegziabher2023patat} could reveal GenAI's advantages over prior systems. System architectures addressing context length limitations, like assessing chunking strategies' efficacy \cite{chen2023intent}, also require more investigation.
\section{Conclusion}

This study investigated the collaborative experience of qualitative researchers using the leading GenAI application, ChatGPT, for thematic analysis. Our findings indicate that researchers value ChatGPT's potential to enhance efficiency and deepen qualitative data comprehension. However, concerns exist regarding trust assurance mechanisms and the practicality of the current interface for interactive analysis. Drawing from these insights, we synthesized five design recommendations that aim to inform the development of future GenAI systems that not only facilitate seamless human-AI collaboration in qualitative research but also enhance user trust and confidence in the technology.

\begin{acks}
This research was at least in part funded by the Australian Research Council (DP210100060) and Jacobs Foundation (Research Fellowship).
\end{acks}

\bibliographystyle{ACM-Reference-Format}
\bibliography{0_reference.bib}






\end{document}